\documentclass[letterpaper]{article} 
\usepackage{aaai2026}  
\usepackage{times}  
\usepackage{helvet}  
\usepackage{courier}  
\usepackage[hyphens]{url}  
\usepackage{graphicx} 
\usepackage{natbib}  
\usepackage{caption} 
\usepackage{algorithm}

\usepackage{subcaption}%
\usepackage{multirow}%
\usepackage{amsmath,amssymb,amsfonts}%
\usepackage{amsthm}%
\usepackage{mathrsfs}%
\usepackage[title]{appendix}%
\usepackage{xcolor}%
\usepackage{textcomp}%
\usepackage{manyfoot}%
\usepackage{booktabs}%
\usepackage{algorithmicx}%
\usepackage{algpseudocode}%
\usepackage{tabularx}

\usepackage{newfloat}
\usepackage{listings}
\DeclareCaptionStyle{ruled}{labelfont=normalfont,labelsep=colon,strut=off} 
\floatstyle{ruled}
\newfloat{listing}{tb}{lst}{}
\floatname{listing}{Listing}
\title{When AI Enters the Workplace, Who Faces Greater Risks? A Gendered Analysis}
\author {
    Miriam Fernandez\textsuperscript{\rm 1},
    Ángel Pavón Pérez\textsuperscript{\rm 1},
    Damiano Giallongo\textsuperscript{\rm 2},
    Davide Ghia\textsuperscript{\rm 2},
    Maryam Yaqub\textsuperscript{\rm 4},
    Daniele Quercia\textsuperscript{\rm 2, 3},
    Tania Cerquitelli\textsuperscript{\rm 2}    
}
\affiliations {
    \textsuperscript{\rm 1}Centre for Protecting Women Online, Open University (UK) \\
    \textsuperscript{\rm 2}Politecnico di Torino (Italy) \\ 
    \textsuperscript{\rm 3}Nokia Bell Labs (UK) \\
    \textsuperscript{\rm 4}Fawcett Society (UK)\\
 }

\usepackage{bibentry}

\begin{document}

\maketitle

\begin{abstract}
Gender inequality remains a persistent structural feature of the labour market, shaping women’s lifetime earnings and economic security. As artificial intelligence (AI) transforms organisational practices, there is growing concern that existing disparities may be unintentionally amplified through task automation, unequal access to upskilling opportunities, and differential returns obtained from technological change. In this paper, we examine how exposure to AI-driven innovation varies across male- and female-dominated occupations, with particular attention to differences across the skill and wage distribution. Using a novel dataset that links occupational characteristics to measures of AI exposure, we analyse how recent advances in Large Language Models (LLMs) and broader AI technologies are distributed across the labour market. Our findings show that, while AI exposure is generally concentrated in higher-skilled and higher-paid occupations for male-dominated occupations, female-dominated occupations display relatively uniform levels of exposure across both high-skilled, high-paid, and low-skilled, low-paid occupations. Moreover, we find that LLM-related exposure is higher in female-dominated occupations, while exposure to broader AI innovation remains more concentrated in male-dominated occupations. A triangulation of these results with existing literature suggests that women, particularly those in the most vulnerable positions (lower-skilled and lower-paid female-dominated occupations), may face greater exposure to forms of AI associated with task automation, job restructuring, reduction of wages and limited career progression. 

\end{abstract}

 \begin{links}
\link{Project page}{https://social-dynamics.net/ai-impact/gender/}
 \end{links}

\section{Introduction}
\label{sec:introduction}

Inequalities in labour markets, such as occupational segregation (i.e., the unequal distribution of men and women across different occupations and industries)\cite{eige2025work, ec_women_labour_market}, the under-representation of women in higher-paying and senior roles \cite{cook2024csuitegap}, and the effects associated with motherhood and caring responsibilities \cite{ons2025motherhood}, remain among the most persistent and structural forms of gender discrimination, shaping women’s lifetime earnings, career progression, and long-term economic security. Despite sustained policy attention, these disparities continue to characterise labour markets across the world.

As AI increasingly transforms organisational practices, there is a growing risk that existing inequalities may be unintentionally reinforced. This may occur through several mechanisms, including the automation of tasks disproportionately performed by women, unequal access to reskilling and upskilling opportunities, or biases embedded in data-driven and algorithmic decision-making systems used for hiring, performance monitoring or promotions \cite{fawcett_ai_gender_2026}.

Existing research on the gender-differentiated effects of AI has primarily focused on identifying the AI exposure of different occupations \cite{WEF2025_future_jobs, ILO2025_women_work_g20, ilo2026genai, jsa2025genai} and whether those occupations are male or female-dominated. However, less is known about how different forms of AI exposure are distributed across the wage and skill structure of male- and female-dominated occupations, and, crucially, how these patterns map onto inequality. Lower-paid and lower-skilled workers (who typically have weaker bargaining power, more limited access to training, and greater employment insecurity) may be disproportionately exposed to AI-related automation and restructuring. Breaking down exposure at this level is therefore essential to identify impact within the most vulnerable groups.

In this paper, we examine how exposure to AI-driven innovation varies across male- and female-dominated occupations, with particular attention to differences across the skill and wage distribution. To investigate these issues, we construct a novel dataset linking occupational characteristics with measures of AI exposure. Specifically, we combine O*NET occupational data (see Section \ref{subsec:ONET}) with U.S. Census data on gender-disaggregated income distributions (see Section \ref{subsec:US_census_data}). Unlike prior studies relying on a single measure of AI exposure, we combine two complementary indices that capture distinct technological paradigms: the Anthropic Index (see Section \ref{subsec:anthropic_index}), which captures Large Language Models (LLM)-related exposure associated with recent advances in generative AI, and the Artificial Intelligence Index (AII) (see Section \ref{subsec:AII_index}) \cite{Septiandri2022}, which captures exposure to broader AI innovation (e.g. robotics) based on patent data.

Our results show that while AI exposure for male-dominated occupations is often associated with higher-skilled and higher-paid occupations, this relationship does not hold within female-dominated occupations. Instead, AI exposure in these occupations is distributed more evenly across the skill and wage spectrum. Moreover, we find that LLM-related exposure is higher in female-dominated occupations, while exposure to broader AI innovation remains more concentrated in male-dominated occupations.

It's important to note that AI exposure does not necessarily imply realised automation or displacement effects. Rather, exposure measures capture the extent to which occupational tasks are likely to be affected by AI systems, whether through augmentation, restructuring, or automation. 

Therefore, to interpret our findings, we triangulate our empirical results with insights from the existing literature on AI and labour markets. This combined analysis suggests that, while men are more likely to benefit from AI as a complement to high-skilled work, women, particularly those in lower-skilled and lower-paid, female-dominated occupations, are more exposed to forms of AI associated with task automation and job restructuring. In these occupations, AI exposure is less likely to translate into productivity-enhancing augmentation and more likely to involve the reorganisation or substitution of tasks, increasing the risk of reduced wages, limited career progression and job displacement. Existing evidence also suggests that participation in adult learning and training is substantially lower among low-skilled workers \cite{oecd2018jobs}, reducing opportunities for upskilling and adaptation to technological change. Taken together, these patterns suggest that AI adoption may generate uneven labour market effects, disproportionately exposing women in more vulnerable positions to displacement risks, reduced wages, and constrained career advancement.


The remainder of the paper is structured as follows. Section \ref{sec:literature} reviews related literature. Section \ref{sec:experimental_setup} describes the experimental setup. Section \ref{sec:data_integration} describes the data integration methodology. Section \ref{sec:US_data_analysis} presents the empirical results. Section \ref{sec:discussion} discusses the work and limitations, and Section \ref{sec:conclusions} concludes.

\section{Literature Review}
\label{sec:literature_review}

A growing body of literature examines how technological change, digitalisation, and more recently AI, may affect gender disparities in the labour market \cite{fawcett_ai_gender_2026}. While technological innovation can increase productivity and create new opportunities, existing evidence suggests that unequal access to technology, differences in adoption, and occupational segregation may widen existing gender gaps in wages, employment, and career progression.

\textbf{Gender differences in the adoption of AI and digital technologies: } Recent studies show consistent gender differences in the adoption and use of emerging technologies, particularly LLMs. Using U.S. survey data, \citet{aldasoro2024gen} find that men report substantially higher use of generative AI tools than women (50\% versus 37\%). The authors identify differences in knowledge about the technology as the primary driver of this gap, accounting for roughly three-quarters of the difference, with the remainder explained by gender differences in trust and privacy concerns. Women, on average, report higher perceived risks and lower levels of trust in AI systems.


Cross-country evidence confirms that these patterns are widespread. \citet{otis2024global}, analysing results from 18 studies covering more than 140,000 individuals globally, found that women consistently adopt generative AI tools at lower rates across most regions, sectors, and occupations. The authors warn that persistent differences in adoption may create a self-reinforcing cycle. Lower participation of women in AI use may lead to systems trained on data that insufficiently reflect women’s needs and preferences, potentially reinforcing existing inequalities in economic outcomes.

\citet{tang2025gender} show how differences in usage translate into differences in productivity gains. Male researchers use generative AI tools more frequently and experience larger efficiency improvements from their use. Such differences in productivity gains may, over time, contribute to widening performance-based pay differences and career progression.

\textbf{Unequal access to high-paying technology-intensive roles:} Beyond differences in adoption, several studies highlight unequal access to technology-intensive jobs as an important mechanism through which innovation may affect gender gaps. \citet{tambe2025hidden} show that women are less likely to apply for jobs requiring expertise in newly emerging technologies, even when controlling for qualifications. Because these positions tend to offer higher wages and stronger career progression, women’s under-representation in such roles contributes to widening pay differences within the IT sector. These findings are consistent with broader evidence on occupational segregation, which shows that women remain under-represented in technical fields \cite{wef_gender_gap_2023}. As technological change often increases returns to specialised skills, unequal participation in these occupations may translate directly into unequal wage growth.

\textbf{AI exposure, automation and wage inequality:} Empirical evidence on the labour-market effects of AI exposure provides mixed results, but several studies suggest that technological change can increase gender disparities under certain conditions. Joint research from the Organisation for Economic Co-operation and Development (OECD) and the International Labour Organisation (ILO) of G20 labour markets shows that gender wage gaps and participation gaps remain substantial despite technological progress, suggesting that innovation alone does not reduce inequality \cite{ILO2025_women_work_g20}.

\citet{pavlenkova2024effects}, analysing administrative data from Estonia between 2006 and 2018, found that the introduction of automation was associated with an increase in the gender pay gap. The authors attribute this effect to differences in occupational structure, with women more likely to work in roles where automation reduced wages or limited career progression. Similar results have been described by the World Economic Forum (WEF), the International Labour Organisation (ILO) and the Australian government. A 2025 WEF report \cite{WEF2025_future_jobs} indicates that automation and AI could affect a substantial share of roles in administrative and service sectors by 2030. These occupations are disproportionately held by women. ILO \cite{ILO2023_generative_ai_jobs, ilo2026genai} also finds that female-dominated occupations face higher exposure to Gen AI-related disruption than male-dominated ones, reflecting the combined effects of occupational segregation and task composition. A recent paper by the Australian government \cite{jsa2025genai} also reports that Gen AI is set to automate routine clerical and administrative work while serving as a powerful assistant for high-skilled roles.


\textbf{Biases in AI systems used for recruitment and performance monitoring:} Policy reports increasingly highlight the risk that AI systems may embed or amplify existing social inequalities if they are developed using biased data or deployed in unequal labour-market contexts. OECD and The Global Partnership on Artificial Intelligence (GPAI) \cite{GPAI2024_substantive_equality_ai} note that AI systems trained on historical data may reproduce existing gender inequalities in hiring, promotion, and pay, unless explicit mitigation strategies are implemented. By conducting a literature analysis, Yaqub \cite{fawcett_ai_gender_2026} also highlights the fairness risks associated with AI systems used for recruitment, performance monitoring, promotions and career progression. 

\textbf{AI systems for decision support:} AI systems are increasingly used in decision support contexts that relate to pay and progression. LLMs are now used to assist with salary benchmarking, negotiation preparation, performance feedback, and career planning. Emerging evidence suggests that these systems can produce different outputs depending on gendered inputs. Experimental testing has shown that identical prompts relating to salary negotiation can yield different recommendations depending on the perceived gender of the user. In a 2025 study \cite{sorokovikova2025surface}, across multiple tested AI models, showed how women were consistently advised to seek lower salaries. While large language models may appear neutral, they can reproduce underlying structural patterns when generating evaluative or advisory outputs.

\textbf{AI exposure across skill and wage distribution within male- and female-dominated occupations:} A key innovation of this paper is that it moves beyond existing studies that examine AI exposure in isolation and investigates whether AI exposure differs systematically across the skill and wage distribution within male- and female-dominated occupations. Our goal is to study how AI will affect the most vulnerable sectors of the population, i.e., those with lower skills and lower pay. Empirically, our findings reveal an important asymmetry that is largely absent from the current literature: AI exposure in male-dominated occupations is concentrated in higher-skilled and higher-paid roles, whereas exposure within female-dominated occupations is distributed more evenly across the wage and skill spectrum, including lower-paid and lower-skilled work. This not only means that AI is likely to affect female-dominated occupations, as previous literature has highlighted, but that it will affect women in the most vulnerable positions.

\label{sec:literature}

\section{Experimental Set-Up}
\label{sec:experimental_setup}

In this section, we describe the datasets used for experimentation as well as the indices selected to quantify AI exposure across occupations.

\subsection{O*NET}
\label{subsec:ONET}

Occupational characteristics are obtained from the ONET database \cite{onet2023}, maintained by the U.S. Department of Labour. ONET provides detailed information on the content of work for nearly 1,000 occupations classified according to the US Standard Occupational Classification (SOC code) system. For each occupation, the database reports a set of workplace tasks describing the activities typically performed on the occupation, together with an importance score that reflects their relevance within the occupation. Tasks are categorised as \emph{core} or \emph{supplemental} depending on their importance and relevance for a given occupation. In addition to task information, the dataset provides quantitative measures of preparation, skills, knowledge, abilities, and work context. These indicators are constructed from surveys of both workers and occupational experts and are designed to capture the requirements and characteristics of each occupation. 
Table \ref{tab:ex_occupation} shows examples of occupations from different job zones extracted from the O*NET database. Job zones (5 to 1) reflect the level of skills require for an occupation.  

\begin{table*}[h]
    \centering
    \tiny
    \begin{tabular}{p{0.18\textwidth}|p{0.2\textwidth}|p{0.05\textwidth}|p{0.4\textwidth}}
    \toprule
        Sector & Occupation & Job Zone & Example Task\\
    \midrule
        Health care and social assistance & Orthodontists & 5 & Provide patients with proposed treatment plans and cost estimates.\\
        Information & Software Developers & 4 & Develop or direct software system testing or validation procedures.\\
        Construction & Electricians & 3 & Disassemble defective electrical equipment, replace defective or worn parts, and reassemble equipment, using hand tools.\\
        Manufacturing & Industrial Truck and Tractor Operators & 2 & Manually or mechanically load or unload materials from pallets, skids, platforms, cars, lifting devices, or other transport vehicles. \\
        Accommodation and food services & Waiters and Waitresses & 1 & Collect payments from customers.\\
    \bottomrule
    \end{tabular}
    \caption{Examples of O*NET occupations with corresponding sector, job zone, and task.}
    \label{tab:ex_occupation}
\end{table*}

\subsection{US Census Data}
\label{subsec:US_census_data}

Occupational and income data for the United States are sourced from the American Community Survey (ACS), a survey conducted by the U.S. Census Bureau \cite{uscensus2024}. The ACS provides detailed annual estimates on the economic and social characteristics of the U.S. population. For this study, we utilise the 1-year estimates from 2024, which represent the latest available 1-year data.

The dataset provides insights into the workforce through specific 'B-series' tables that classify data using Census Occupational Classification (OCC) codes, which are aligned with and derived from the US Standard Occupational Classification (SOC) system. We utilise tables \textit{B24124}, \textit{B24125}, and \textit{B24126} to obtain the total counts of individuals in detailed occupations for the full-time year-round male and female populations, respectively. These counts allow for the calculation of gendered occupational density. Furthermore, we incorporate median earnings data from tables \textit{B24121} (total median earnings), \textit{B24122} (male median earnings), and \textit{B24123} (female median earnings). In table \ref{tab:ussample}, we show a sample from the US census data. 

\begin{table}[h]
\tiny
\begin{tabular}{@{}lcccc@{}}
\toprule
Occupation         & \begin{tabular}[c]{@{}l@{}}Estimated \\ number \\ of males\end{tabular} & \begin{tabular}[c]{@{}l@{}}Estimated \\ number \\ of females\end{tabular} & \begin{tabular}[c]{@{}l@{}}Estimated Median \\ Earnings Males\end{tabular} & \begin{tabular}[c]{@{}l@{}}Estimated Median \\ Earnings Females\end{tabular} \\ \midrule
Chief executives   & 1,018,958                                                               & 434,992                                                                   & 191,756\$                                                                    & 151,010\$                                                                      \\
Marketing managers & 216,507                                                                 & 355,414                                                                   & 106,963\$                                                                    & 93,869\$                                                                       \\
Financial managers & 632,871                                                                 & 774,201                                                                   & 124,972\$                                                                    & 82,680\$                                                                       \\ \bottomrule
\end{tabular}
\caption{A sample from U.S. Census data, extracted from tables B24125, B24126, B24122, and B24123, showing occupations (U.S. SOC descriptions), the estimated number of males and females, and the estimated median annual earnings by gender.}
\label{tab:ussample}
\end{table}

\subsection{AII Index}
\label{subsec:AII_index}

The AI Impact Index (AII) \cite{Septiandri2022} is an occupation-level measure of exposure to Artificial Intelligence. It is defined as the proportion of tasks within an occupation that are affected by AI. To determine whether a task is exposed to AI, the original study computes the cosine similarity between textual descriptions of AI-related patents and occupational task descriptions. A task is considered exposed if its highest similarity score with any AI patent exceeds the threshold corresponding to the 90th percentile of the similarity distribution. Occupational task structures and task descriptions are obtained from the O*NET database \cite{onet2023}.

Patented software and applications often require several years before being implemented in real-world work environments. Therefore, the AII index does not necessarily reflect current occupational shifts but rather potential future shifts.  Moreover, LLMs, currently among the most widely recognised and utilised AI technologies, are frequently not patented \cite{openai2025gpt5, openai2024gpt4technicalreport, bai2022Claude}. Some are released as open-source systems, while others remain proprietary. As a result, relying on patents to measure AI exposure does not necessarily capture the immediate impact of LLMs on occupations; instead, it provides an indication of broader AI impact. To capture how different occupations are exposed to LLM-related innovation, we use the Anthropic Index, explained below.


\subsection{Anthropic Index}
\label{subsec:anthropic_index}

In March 2026, Anthropic released a new report (\emph{Labor market impacts of AI: A new measure and early evidence}) and with it, a new occupation-level AI exposure index. The Anthropic Index corresponds to the occupation-level \texttt{observed\_exposure} score released by Anthropic \cite{massenkoffmccrory2026labor,massenkoffmccrory2026appendix,anthropic2026labor_market_impacts_dataset}. 

This score is constructed from the previous two Anthropic Economic Index datasets, covering Claude usage data from August 2025 and November 2025. These data comprise 2 million Claude.ai observations and 2 million first-party API observations, and task coverage is defined using a minimum threshold of 100 work-related uses, corresponding to 0.0025\% of total traffic \cite{massenkoffmccrory2026appendix,massenkoffmccrory2026labor}.

\section{Data Integration} 
\label{sec:data_integration}
In this section, we describe the process of data integration to obtain gendered disaggregated information for different occupations.   

Data from the U.S. Census Bureau are classified according to the Census Occupational Classification (OCC) system, which constitutes the Bureau’s own occupational coding scheme. To ensure comparability with other occupational datasets, the U.S. Census Bureau provides an official crosswalk that maps Standard Occupational Classification (SOC) codes to OCC codes. In contrast, the O*NET database identifies occupations using the O*NET-SOC coding system. This classification is more granular than the standard SOC system, as it splits several SOC occupations into more detailed categories, resulting in a larger number of occupational entries. Since both AI exposure indices are constructed using the O*NET classification, two mapping steps are required to align them with Census data: (1) from O*NET-SOC codes to SOC codes and (2) from SOC codes to OCC codes. 

\subsection{AII Index}
\label{subsec:data_integration_US_AII}
The AII index is available for 873 out of the 923 occupations identified in the O*NET classification. When mapping O*NET-SOC codes to U.S. Census OCC codes, 517 out of 570 OCC occupations are successfully matched to a corresponding SOC code with an associated AII value. The remaining 53 occupations do not have AII data available.

Because some OCC codes map to multiple SOC codes, the corresponding AII value is computed as the Geometric Exposure Index (see Section \ref{subsec:gei}) of the AII scores of the matched SOC occupations. Dispersion in AII values may arise when aggregating across classifications. To limit aggregation noise and ensure consistency across mappings, we exclude occupations for which the standard deviation of the associated AII values exceeds a predefined threshold ($\sigma > 0.15$). This value represents a moderate dispersion threshold for an index bounded between 0 and 1. Results are robust to alternative cutoffs (e.g., 0.10 or 0.20). This procedure results in the exclusion of five occupations from the sample. Occupations with missing values in relevant columns (e.g., number of female workers) were filtered out. 


The final dataset, combining U.S. Census data with the AII index, includes
505 occupations. The AII series in Figure~\ref{fig:sector_coverage} shows broad
sectoral coverage, with 86.1\% of detailed Census occupations preserved in the
filtered occupation-level dataset. Coverage exceeds 80\% in nearly all sectors
and reaches 100\% in several cases. The only notable shortfall is Transportation
(63.6\%), while Military Specific is entirely absent, likely due to limited
underlying data availability.

    

\subsection{Anthropic Index}
\label{subsec:data_integration_US_Anthropic}

The Anthropic Index is matched to US census occupations using the occupation-level \texttt{observed\_exposure} score described above. It is available for 756 out of the 923 occupations identified in the O*NET classification. When mapping O*NET-SOC codes to U.S. Census OCC codes, 414 out of 570 OCC occupations are successfully matched to a corresponding SOC code with an associated Anthropic Index value. The remaining 156 occupations do not have Anthropic data available.

Because some OCC codes map to multiple SOC codes, the corresponding Anthropic Index value is computed as the Geometric Exposure Index (see Section \ref{subsec:gei}) of the Anthropic Index scores of the matched SOC occupations.



\begin{figure*}[t]
    \centering
    \includegraphics[width=1\textwidth]{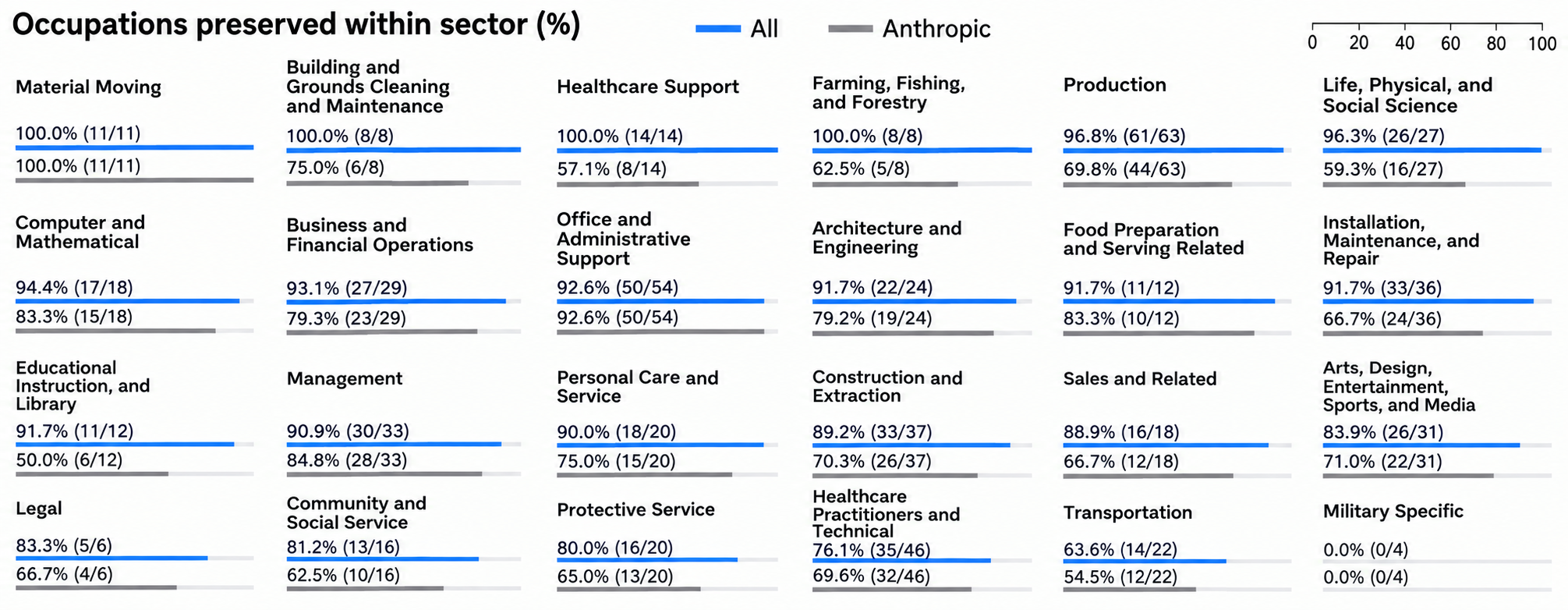}
    \caption{Coverage of sector-classified detailed U.S. Census occupations
    in the AII and Anthropic occupation-level datasets. For each sector, the
    upper blue bar reports AII coverage and the lower gray bar reports
    Anthropic coverage.}
    \label{fig:sector_coverage}
\end{figure*}

Our final analytical sample retains 394 out of 570 U.S. Census occupations,
as gender-disaggregated data are missing for 20 of the matched occupations.
The Anthropic series in Figure~\ref{fig:sector_coverage} shows that sectoral
coverage is at least 50\% in every sector except Military Specific, which is
entirely absent.


\subsection{Geometric Exposure Index}
\label{subsec:gei}

As discussed in the previous sections, some OCC codes correspond to multiple SOC codes. To assign a single exposure value to each OCC code, we aggregate the corresponding SOC exposure values. Specifically, we compute a unique AII value and a unique Anthropic Index value for each OCC code using the Geometric Exposure Index (GEI).

The distributions of both AII and the Anthropic Index are right-skewed, exhibiting long tails. Because they are not normally distributed, the mean is not a reliable summary statistic. While the geometric mean is better suited to skewed distributions, it introduces another issue: it requires all values to be strictly positive. In both AII and the Anthropic Index, however, zero values occur, representing occupations with no exposure to AI. To address this limitation, we propose the Geometric Exposure Index, which integrates the robustness of the geometric mean with the interpretability of the expected value.

\textbf{Prevalence:}
\begin{equation}
p=P(AI\_Index>0)=1-P(AI\_Index=0)
\label{eq:gei_prevalence}
\end{equation}
where $AI\_Index$ is the index considered (Anthropic or AII), and $P$ is the probability, i.e. the frequency, of $AI\_Index>0$.

\textbf{Conditioned Intensity:} geometric mean on positive values only.
\begin{equation}
GM_+=\exp\!\bigl(\mathrm{mean}(\log(AI\_Index)\mid AI\_Index>0)\bigr)
\label{eq:gei_gm_positive}
\end{equation}

\textbf{Geometric Exposure Index:}
\begin{equation}
GEI=p \cdot GM_+
\label{eq:gei_definition}
\end{equation}

The GEI can be interpreted as an expected value:
\begin{equation}
E[AI\_Index]=p \cdot E[AI\_Index \mid AI\_Index>0]
\label{eq:gei_expected_value}
\end{equation}
where $E[AI\_Index \mid AI\_Index>0]$ is the mean of positive values.

\section{Data Analysis}
\label{sec:US_data_analysis}
In this section, we present the analyses conducted using the integrated dataset. We examine two main relationships:
(i) the association between occupational gender composition and exposure to AI, and (ii) the relationship between occupational skill and wage levels and exposure to AI. 


\subsection{Occupational Gender Composition and AI Exposure}
\label{subsec:US_occupational_gender_composition_and_AI_exposure}

In our first analysis, our aim was to observe whether occupations with a higher number of women (female-dominated occupations) would be more or less exposed to AI than occupations with a higher number of males (male-dominated occupations). 

Table \ref{tab:top_occupations_female_share_male_dom} presents examples of occupations with a high male share. These include bus and truck mechanics and diesel engine specialists, heating, air conditioning, and refrigeration mechanics and installers, as well as electric motor, power tool, and related repairers, among others. These occupations are typically associated with technical or manual skill requirements and have historically shown low female participation. Examples of occupations with a high female share (see Table \ref{tab:top_occupations_female_share_female_dom}) include nurse midwives, skin care specialists, speech-language pathologists, legal secretaries and administrative assistants, and childcare workers. These occupations are more frequently concentrated in care, administrative, and service-related activities, which have traditionally employed a larger proportion of women.


The distribution shown in these tables highlights the persistent occupational segregation in the labour market, with men and women concentrated in different types of roles. 

\begin{table}[htbp]
    \centering
    \tiny
    \setlength{\tabcolsep}{2.5pt}   
    
    \begin{tabular}{r p{3.0cm} c r r r}
        \toprule
        Rank & Occupation & SOC Code & \% Females & \% Males & No. Workers \\
        \midrule
        1  & Elevator and escalator installers and repairers & 47-4021 & 0.6 & 99.4 & 27775 (0.03\%) \\
        2  & Earth drillers, except oil and gas & 47-5023 & 1.3 & 98.7 & 23261 (0.03\%) \\
        3  & Automotive glass installers and repairers & 49-3022 & 1.5 & 98.5 & 13874 (0.02\%) \\
        4  & Crane and tower operators & 53-7021 & 1.6 & 98.4 & 55344 (0.07\%) \\
        5  & Bus and truck mechanics and diesel engine specialists & 49-3031 & 1.6 & 98.4 & 262702 (0.32\%) \\
        6  & Plasterers and stucco masons & 47-2161 & 1.6 & 98.4 & 15444 (0.02\%) \\
        7  & Heating, air conditioning, and refrigeration mechanics and installers & 49-9021 & 1.7 & 98.3 & 440629 (0.53\%) \\
        8  & Electric motor, power tool, and related repairers & 49-2092 & 1.8 & 98.2 & 18403 (0.02\%) \\
        9  & Millwrights & 49-9044 & 1.8 & 98.2 & 41668 (0.05\%) \\
        10 & Plumbers, pipefitters, and steamfitters & 47-2152 & 1.8 & 98.2 & 521656 (0.63\%) \\
        \bottomrule
    \end{tabular}
    \caption{Occupations (having 0--39\% females) ranked by \% males. Values in parentheses in the No. Workers column report the percentage of all workers in the dataset present in each occupation.}  \label{tab:top_occupations_female_share_male_dom}
\end{table}


\begin{table}[htbp]
    \centering
    \tiny
    \setlength{\tabcolsep}{2.5pt}

    \begin{tabular}{r p{3.0cm} c r r r}
        \toprule
        Rank & Occupation & SOC Code & \% Females & \% Males & No. Workers \\
        \midrule
        1  & Skincare specialists & 39-5094 & 98.3 & 1.7 & 52438 (0.06\%) \\
        2  & Speech-language pathologists & 29-1127 & 95.3 & 4.7 & 139011 (0.17\%) \\
        3  & Legal secretaries and administrative assistants & 43-6012 & 94.9 & 5.1 & 42667 (0.05\%) \\
        4  & Childcare workers & 39-9011 & 94.5 & 5.5 & 402669 (0.49\%) \\
        5  & Executive secretaries and executive administrative assistants & 43-6011 & 93.1 & 6.9 & 230479 (0.28\%) \\
        6  & Medical secretaries and administrative assistants & 43-6013 & 93.0 & 7.0 & 83322 (0.10\%) \\
        7  & Dental assistants & 31-9091 & 92.6 & 7.4 & 197365 (0.24\%) \\
        8  & Secretaries and administrative assistants, except legal, medical, and executive & 43-6014 & 92.5 & 7.5 & 1444384 (1.75\%) \\
        9  & Dental hygienists & 29-1292 & 90.4 & 9.6 & 93601 (0.11\%) \\
        10 & Medical assistants & 31-9092 & 90.3 & 9.7 & 503025 (0.61\%) \\
        \bottomrule
    \end{tabular}
    \caption{Occupations (having 60--100\% females) ranked by \% females. Values in parentheses in the No. Workers column report the percentage of all workers in the dataset represented by each occupation.}    \label{tab:top_occupations_female_share_female_dom}
\end{table}

\begin{figure*}[htbp]    
    \centering
    \begin{subfigure}{0.48\linewidth}
        \centering
        \includegraphics[width=\linewidth]{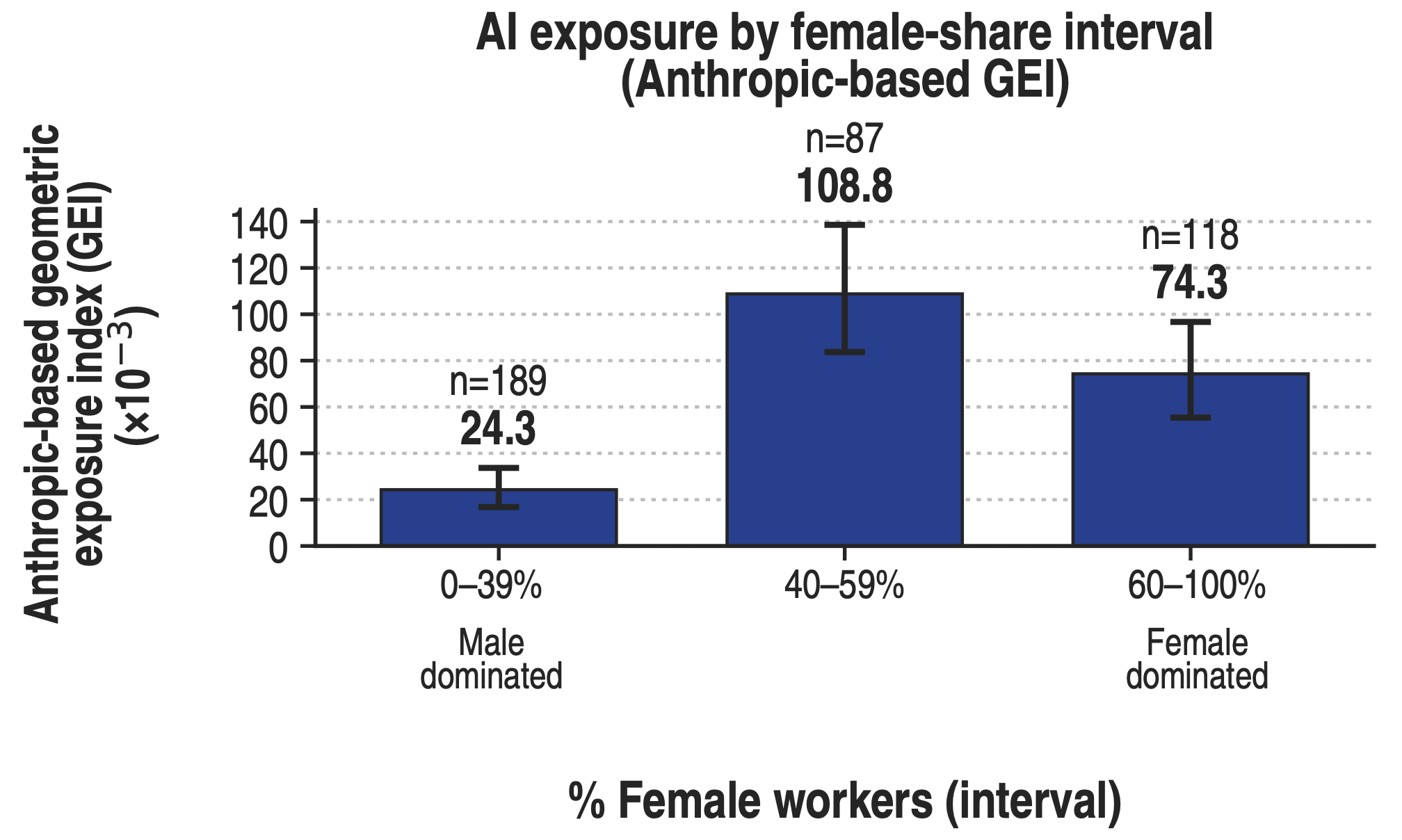}
        \caption{\textbf{Anthropic Index}}
        \label{fig:anthropic_female_share}
    \end{subfigure}
    \begin{subfigure}{0.48\linewidth}
        \centering
        \includegraphics[width=\linewidth]{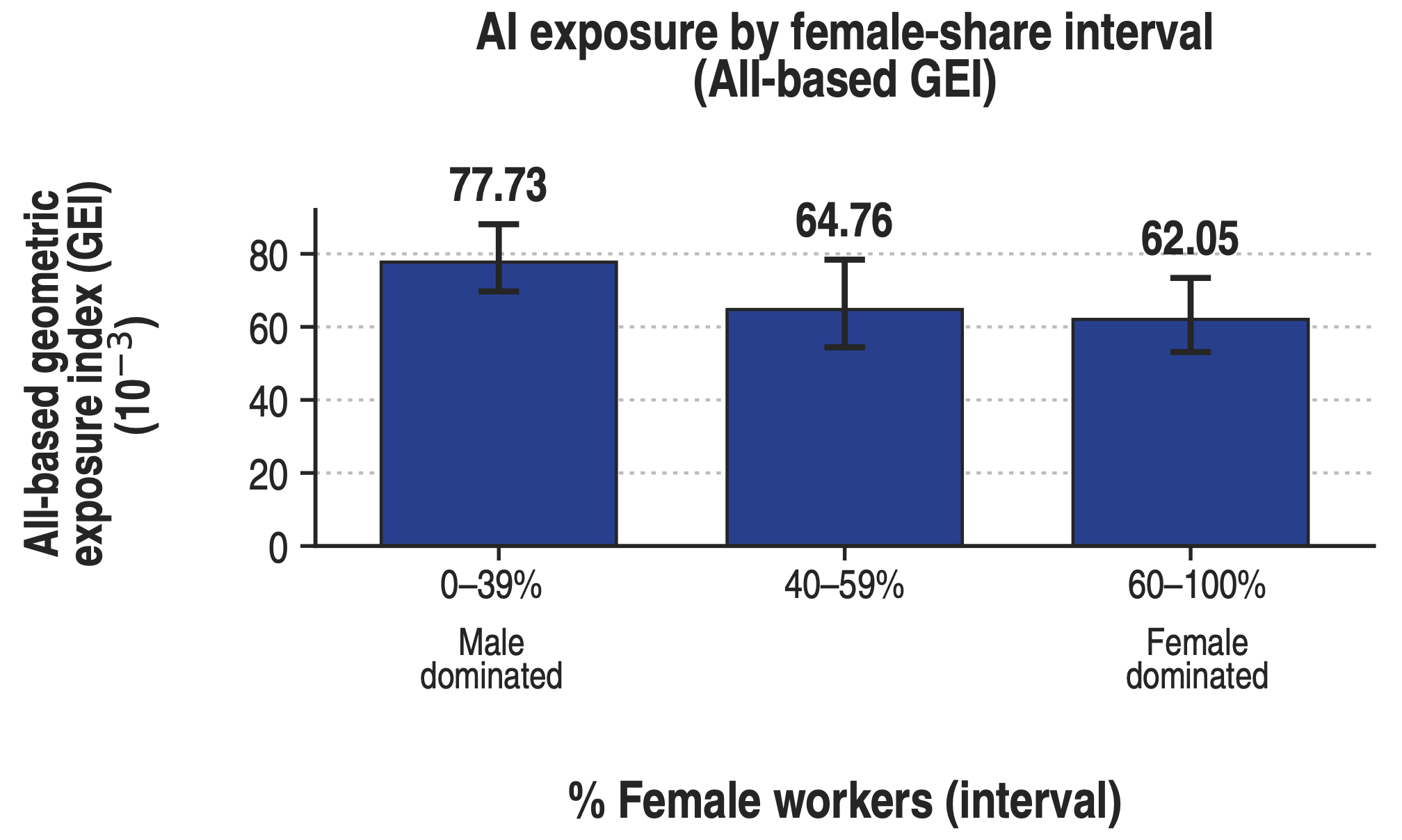}
        \caption{\textbf{AII}}
        \label{fig:aii_female_share}
    \end{subfigure}
    \caption{AI exposure across intervals of the female share of workers. Error bars represent 95\% confidence intervals. Occupations with predominantly female workers (more than 60\%) tend to exhibit higher Anthropic Index values (74.3 \emph{vs.} 24.3) but lower AII values (62.05 \emph{vs.} 77.73) than occupations with relatively few female workers (less than 40\%). 
    }
    \label{fig:female_share}
\end{figure*}

The results of our analysis are presented in Figure \ref{fig:female_share}. For the Anthropic Index (Figure~\ref{fig:anthropic_female_share}), the pattern is non-monotonic. Mixed occupations ($40-59\%$ female) exhibit the highest values, while male-dominated occupations have the lowest, and female-dominated occupations lie in between. ANOVA test confirmed statistically significant difference between distributions ($F=18.39,p<0.001$). Pairwise comparisons through Tukey test show that male-dominated occupations have significantly lower Anthropic Index values than both mixed and female-dominated occupations ($p<0.001$).

In contrast, occupations with a higher female share are associated with lower AII values. Male-dominated occupations exhibit the highest exposure, followed by mixed occupations, with female-dominated occupations having the lowest exposure. The Kruskal-Wallis test confirms a statistically significant difference between groups ($H = 7.99, p < 0.05$), and Dunn’s post-hoc test indicates a significant difference between male-dominated and female-dominated occupations ($p < 0.05$).

This pattern indicates that recent advances in large language models (LLMs), captured by the Anthropic Index, are more strongly associated with occupations with higher female representation, while broader measures of AI innovation, as captured by the AII, remain concentrated in male-dominated occupations. Our literature review shows that the use of LLMs in female-dominated occupations, such as clerical and administrative roles, is associated with higher risks of automation, task restructuring, wage reductions, and limited career progression \cite{jsa2025genai, ILO2023_generative_ai_jobs, WEF2025_future_jobs, pavlenkova2024effects}. 

Even in occupations where LLMs may generate productivity gains, the literature highlights that men tend to experience larger productivity gains from their use \cite{tang2025gender}. This suggests that, even when LLMs augment rather than replace labour, men may capture a disproportionate share of the resulting benefits.

\subsection{Occupational Skill and Wage levels
and AI Exposure}
\label{subsec:US_occupational_skill_and_wage_levels_and_AI_exposure}

In this section, we analyse whether exposure to AI differs across occupations with different skill requirements and wage levels. Specifically, we examine whether high-skill and high-pay occupations are more or less exposed to AI than occupations characterised by lower skill requirements and lower earnings. This analysis is important because the distribution of AI exposure across the wage and skill structure of the labour market may influence future inequality. Understanding this relationship is therefore essential for assessing how AI may affect both overall wage inequality and gender disparities, given the uneven distribution of men and women across occupations with different skill and pay levels.

\subsubsection{SW score}
\label{subsubsec:SW_score}
To measure the level of skill and wage associated with an occupation, we propose the SW score. The O*NET database provides information on the level of preparation required for each occupation. In particular, occupations are classified into five categories, known as Job Zones, which reflect the levels of education, experience, and training typically required to perform the job. To identify the most skilled and well-paid occupations, we constructed a composite measure that accounts for both the level of preparation required (as captured by the Job Zone classification) and the median annual earnings associated with each occupation, obtained from U.S. Census data. The measure, called the SW score, is obtained through sequential steps:

\begin{enumerate}
    \item {\textbf{Earning normalisation:} To limit the influence of extreme values, while allowing differences to be interpreted in proportional terms, earning is first log-transformed. This transformation also improves comparability with the discrete Job Zone measure when constructing the composite SW score: 
\begin{equation}
e_o = \log(\text{earning}_o)
\end{equation}
where $o$ indicates the occupation.

}

\item{\textbf{Standardisation: } The job training and education requirements are based on O*NET's job zones, which classify occupations by the level of education, training, and experience required, ranging from zone 1 (minimal education and training, e.g., dishwashers) to zone 5 (extensive education training, e.g., surgeons). For clarity, we rephrased O*NET’s original ``job zones'' as ``job training and education requirements''.

Earning and skill requirements (job zone) are converted into z-scores:
\begin{align}
z^{\text{earning}}_o &= \frac{e_o - \mu_e}{\sigma_e}, \\
z^{\text{skilled}}_o &= \frac{\text{job training}_o - \mu_s}{\sigma_s},
\end{align}

where $\mu_e$ and $\sigma_e$ denote the mean and standard deviation of earnings, and $\mu_s$ and $\sigma_s$ denote the mean and standard deviation of the job zone skill requirement.
} 

\item{\textbf{Combination:} The SW score is defined as a weighted average of the two standardised components:
\begin{equation}
SW_o = w \, z^{\text{earning}}_o + (1 - w) \, z^{\text{skilled}}_o
\end{equation}

where $w = 0.6$ assigns slightly greater importance to earning. Both components capture complementary dimensions of an occupation, but earnings provide a market-based valuation of skills and are therefore given slightly higher weight.}

\end{enumerate}

\subsubsection{SW Analysis}
\label{subsubsec:SW_analysis}
\begin{figure*}[!htbp]
    \centering
    \begin{subfigure}{0.48\linewidth}
        \centering
        \includegraphics[width=\linewidth]{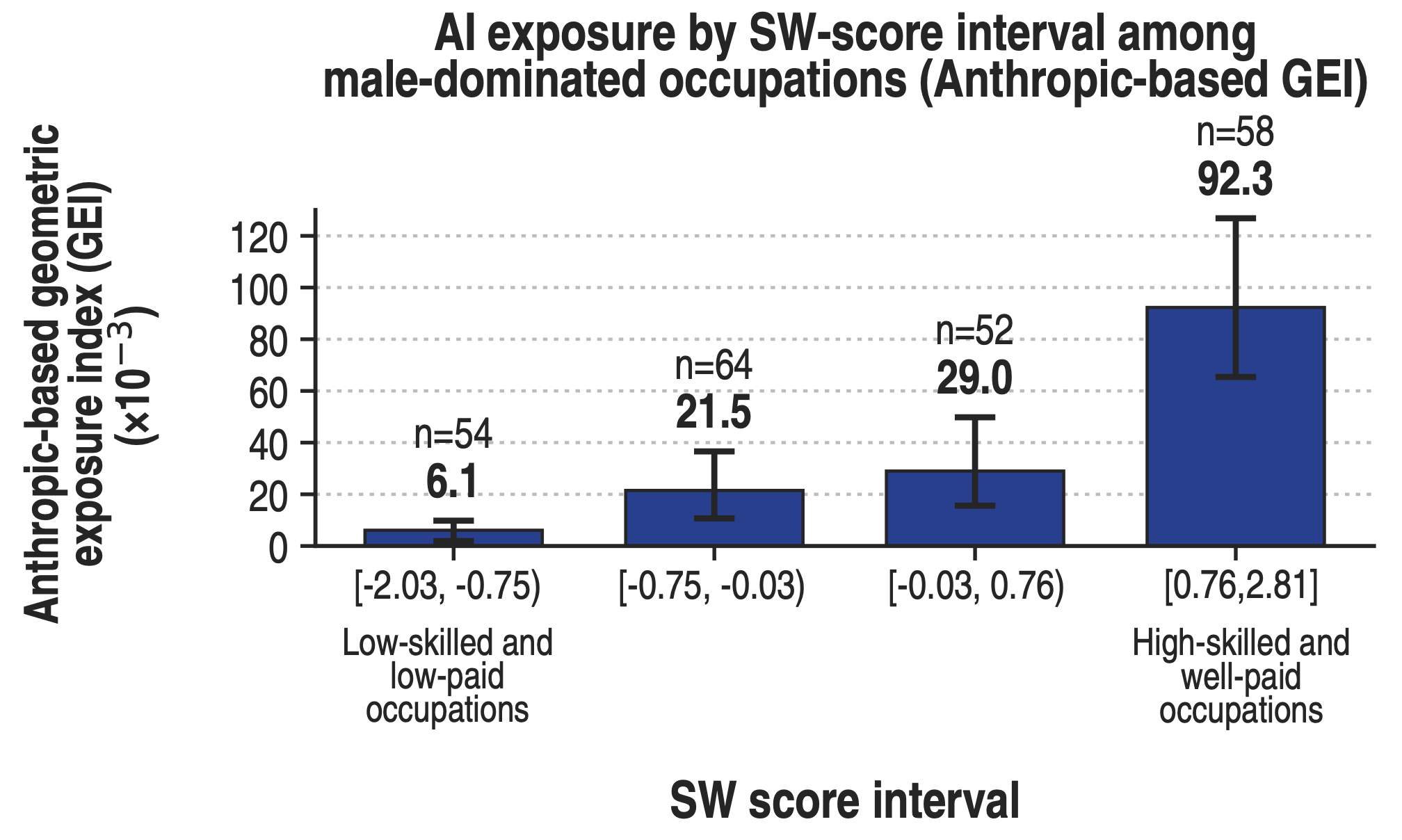}
        \caption{\textbf{Anthropic Index}}
        \label{fig:male_anthropic_sw_score}
    \end{subfigure}
    \begin{subfigure}{0.48\linewidth}
        \centering
        \includegraphics[width=\linewidth]{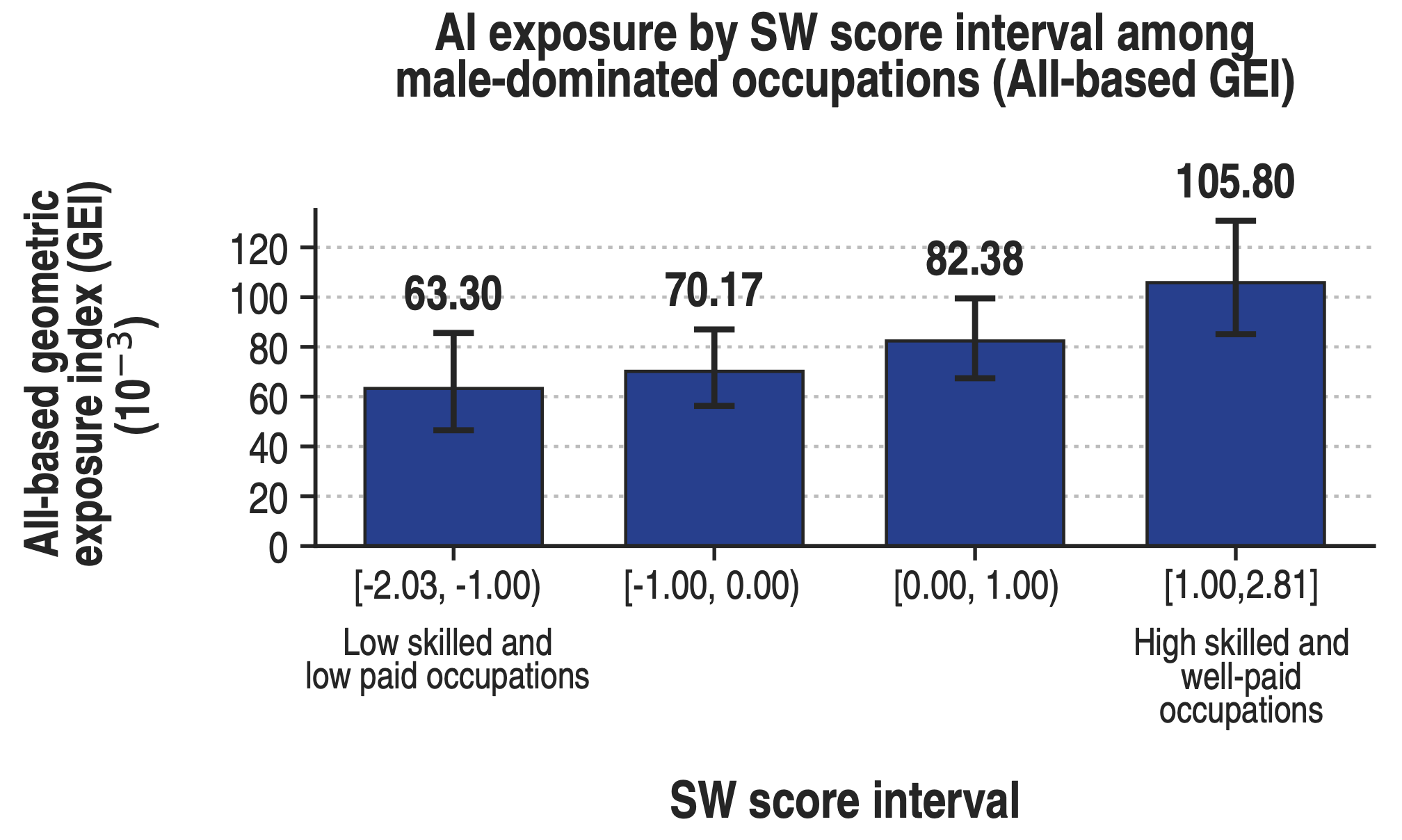}
        \caption{\textbf{AII}}
        \label{fig:male_aii_sw_score}
    \end{subfigure}
    \caption{AI exposure by SW score interval in male-dominated occupations.
    }
    \label{fig:male_exposure_by_sw_score}
\end{figure*}

\begin{figure*}[!htbp]
    \centering
    \begin{subfigure}{0.48\linewidth}
        \centering
        \includegraphics[width=\linewidth]{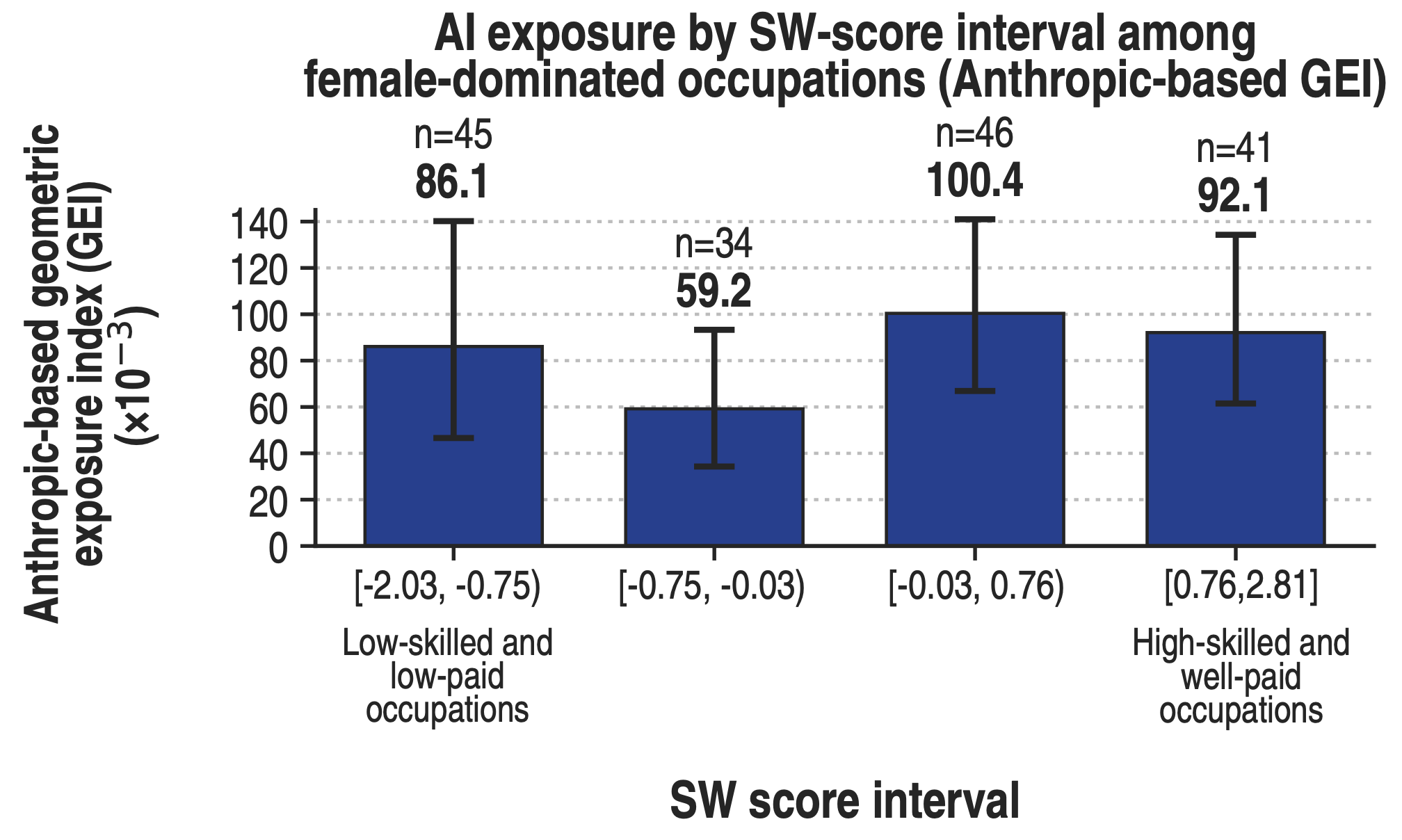}
        \caption{\textbf{Anthropic Index}}
        \label{fig:female_anthropic_sw_score}
    \end{subfigure}
    \begin{subfigure}{0.48\linewidth}
        \centering
        \includegraphics[width=\linewidth]{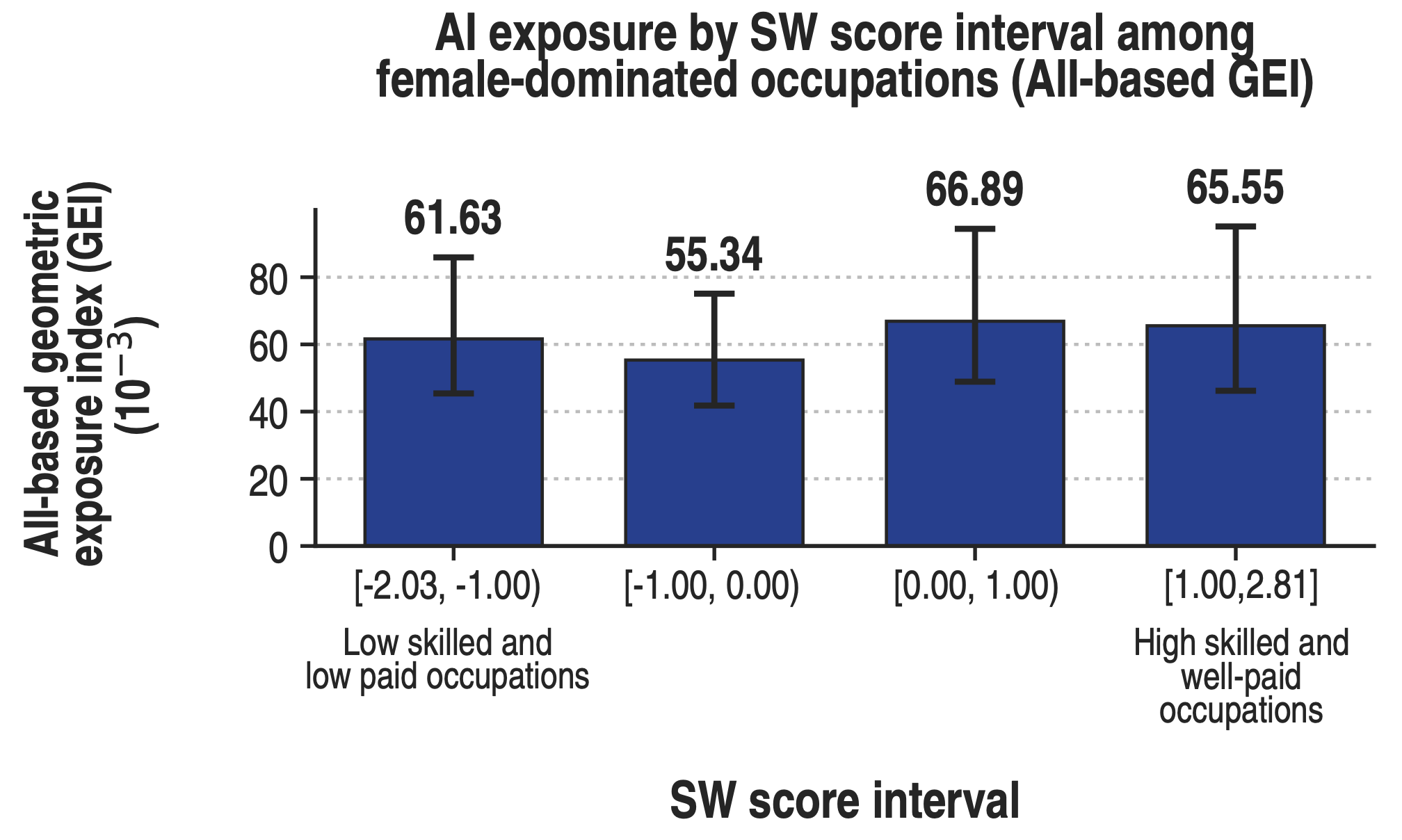}
        \caption{\textbf{AII}}
        \label{fig:female_aii_sw_score}
    \end{subfigure}
    \caption{AI exposure by SW score interval in female-dominated occupations.
    }
    \label{fig:female_exposure_by_sw_score}
\end{figure*}

Using the SW score, we examine how exposure to AI is distributed along the occupational skill and wage structure for both female-dominated (see Figure \ref{fig:female_exposure_by_sw_score}) and male-dominated occupations (see Figure \ref{fig:male_exposure_by_sw_score}). 

For male-dominated occupations, statistically significant differences in exposure are observed across skill and wage levels when using both the Anthropic Index (LLM-based exposure) ($F=16.18,p>0.01$) and AII index ($F=8.35, p<0.05$). This suggests that, within male-dominated occupations, current exposure to LLMs and general AI is not evenly distributed across the skill and wage spectrum, but more concentrated in higher-skilled and higher-paid roles.

For female-dominated occupations, a different pattern emerges. Both Anthropic and AII indexes suggest a more uniform distribution of AI exposure across female-dominated occupations, regardless of skill and wage levels. Statistical tests were not significant for both the indexes ($p>0.05$), confirming the hypothesis. This indicates that AI-related changes will affect a broader range of occupations, including those characterised by lower skill requirements and lower pay.

Our analysis of occupations within the lowest and highest SW bins (see Tables \ref{tab:US_sw_bin1_male_female}, \ref{tab:US_sw_bin2_male_female}, \ref{tab:US_sw_bin3_male_female}, \ref{tab:US_sw_bin4_male_female} for examples) reveals that female-dominated occupations in the lowest SW scores are predominantly concentrated in the administrative and service sectors. To interpret these patterns, we compare the occupational composition of our SW bins with classifications and examples reported in prior studies on AI and automation exposure, including the JSA report \cite{jsa2025genai}, the World Economic Forum Future of Jobs analysis \cite{WEF2025_future_jobs}, and the ILO evidence on generative AI exposure \cite{ILO2023_generative_ai_jobs, gmyrek2025generative}. This comparison is conducted by semi-automatically mapping occupations in our dataset to the occupational groups and exemplar roles discussed in these studies, allowing us to assess whether occupations in the different SW bins correspond to those previously identified as highly exposed to AI-driven automation and/or are considered declining jobs.

\begin{figure}
   \centering
    \includegraphics[width=\columnwidth]{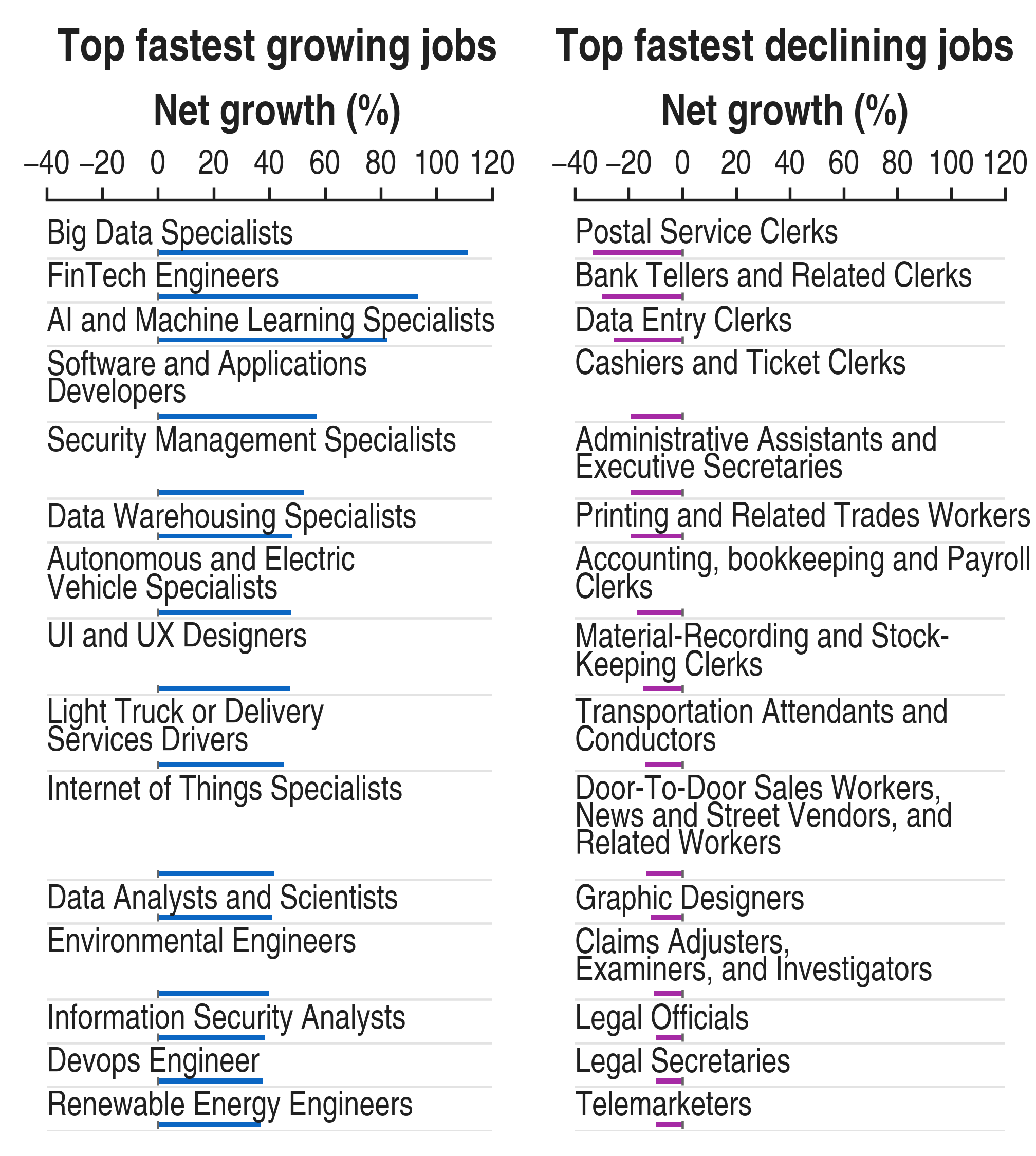}
    \caption{World Economic Forum, Future of Jobs Survey 2024. Image extracted from \cite{WEF2025_future_jobs}. The image has been adapted using Gemini for better visibility.}
    \label{fig:occupations-example}
\end{figure}

First, occupation titles were extracted from the relevant reports and matched against the occupation names in our dataset using automated textual matching techniques, including exact matching and Levenshtein distance. Following this initial mapping stage, the automatically generated matches were manually reviewed to validate semantic correspondence between occupations and to remove incorrect or ambiguous mappings. For occupations that could not be matched through automated textual similarity, additional manual assessment was conducted to identify equivalent or closely related occupational categories based on occupational descriptions and task profiles. This process enabled a comparison between the occupations identified in our analysis and those previously characterised as exposed to AI-driven automation or projected as growing or declining
(see Figure~\ref{fig:occupations-example} for examples of such occupations).

This triangulation shows a strong overlap between low-SW female-dominated occupations in our data and the administrative and service roles that the literature consistently identifies as most susceptible to routine task automation and restructuring. These patterns suggest that women concentrated in these occupations are more likely to face elevated risks of job displacement and task substitution. At the same time, the literature indicates that participation in adult learning and training is significantly lower among low-skilled workers \cite{oecd2018jobs}, which further constrains their ability to adapt through upskilling or reskilling. Taken together, these findings point to the direction that workers in low-SW female-dominated occupations may face compounded vulnerability arising from both higher AI exposure associated with automation and job displacement and weaker access to adaptive labour-market mechanisms.

\section{Discussion}
\label{sec:discussion}

This paper investigates AI exposure across male- and female-dominated occupations, with particular attention to differences across the wage and skill distribution. Our findings suggest that while AI exposure in male-dominated occupations is more strongly concentrated in high-skilled and high-paid occupations, exposure within female-dominated occupations is distributed more evenly across both high-skilled, high-paid and low-skilled, low-paid occupations. This distinction is important because lower-skilled and lower-paid workers are typically more vulnerable to labour-market disruption due to factors such as weaker bargaining power, lower access to training and reskilling opportunities, and greater employment insecurity. Combined with existing evidence from the literature, our results suggest that women in these more vulnerable positions are disproportionately exposed to forms of AI associated with task automation and job restructuring, rather than productivity-enhancing augmentation. As a result, AI adoption may place women in lower-paid and lower-skilled occupations at greater risk of displacement, constrained wages and career progression.

An additional contribution of this work is the distinction between different forms of AI. By combining an LLM-specific exposure index (Anthropic) with a broader AI innovation index (AII), the analysis highlights that different AI technologies may affect workers in different ways. Female-dominated occupations show higher exposure to LLM-related technologies, whereas broader AI innovation remains more concentrated in male-dominated occupations. 

Our findings should, however, be interpreted in light of several limitations. The Anthropic Index is based on a relatively short observation window \cite{anthropic2026labor_market_impacts_dataset}, which may limit its ability to capture stable and representative patterns of LLM use across occupations. This index is constructed from an analysis of Claude conversations, where classifiers map interactions to occupational tasks using the U.S. O*NET taxonomy. This is therefore a usage-based, platform-specific measure of realised LLM interaction. Different LLMs may exhibit different patterns of adoption and use.

Regarding the AII, this index relies on patent data. While patents provide useful insight into the direction of technological development, not all patented innovations are ultimately deployed in practice, and the timing and scale of adoption remain uncertain. 

These two AI exposure indices do not cover identical sets of occupations, which constrains direct comparability between different measures of AI exposure. This limitation is particularly relevant when interpreting differences between LLM-specific exposure and broader AI-related innovation trends. More fundamentally, both indices measure potential exposure to AI rather than realised labour-market outcomes. They do not directly capture whether AI adoption leads to task augmentation, task automation, productivity gains, wage growth, or job displacement. To help interpret the possible implications of different exposure patterns, we triangulate our empirical findings with evidence from the literature. We compare the occupations identified as highly exposed in our analysis with prior empirical studies examining whether AI exposure in similar occupational contexts has been associated with augmentation-oriented outcomes or with automation-related outcomes (e.g., routine task substitution and displacement risks), and/or whether listed occupations are considered growing or declining. 

It is also important to note that our analysis focuses on existing occupations and does not capture the emergence of new occupations or shifts in occupational composition over time. These limitations underline the need for continued empirical research on the long-term relationship between AI exposure and labour-market inequality. Nevertheless, our findings suggest that AI may disproportionately increase risks for women in more vulnerable labour-market positions. This highlights the importance of targeted upskilling and reskilling programmes for women in high-exposure occupations, particularly in administrative and professional services roles, alongside systematic gender-disaggregated monitoring of pay and career progression in AI-intensive sectors to support timely policy intervention.

\begin{table*}[htbp]
    \centering
    \tiny
    \setlength{\tabcolsep}{2.5pt}    
    \resizebox{\textwidth}{!}{%
    \begin{tabular}{r p{6.6cm} c r @{\hspace{0.6cm}} r p{6.6cm} c r}
        \toprule
        \multicolumn{4}{c}{\textbf{Male-dominated occupations}} & \multicolumn{4}{c}{\textbf{Female-dominated occupations}} \\
        \cmidrule(lr){1-4} \cmidrule(lr){5-8}
        Rank & Occupation & SOC Code & SW score & Rank & Occupation & SOC Code & SW score \\
        \midrule
        1  & Dishwashers & 35-9021 & -1.950 & 1  & Hosts and hostesses, restaurant, lounge, and coffee shop & 35-9031 & -2.027 \\
        2  & Dining room and cafeteria attendants and bartender helpers & 35-9011 & -1.814 & 2  & Fast food and counter workers & 35-3023 & -1.961 \\
        3  & Landscaping and groundskeeping workers & 37-3011 & -1.572 & 3  & Laundry and dry-cleaning workers & 51-6011 & -1.810 \\
        4  & Ushers, lobby attendants, and ticket takers & 39-3031 & -1.438 & 4  & Maids and housekeeping cleaners & 37-2012 & -1.802 \\
        5  & Pressers, textile, garment, and related materials & 51-6021 & -1.356 & 5  & Graders and sorters, agricultural products & 45-2041 & -1.788 \\
        6  & Parking attendants & 53-6021 & -1.335 & 6  & Sewing machine operators & 51-6031 & -1.661 \\
        7  & Umpires, referees, and other sports officials & 27-2023 & -1.244 & 7  & Food preparation workers & 35-2021 & -1.457 \\
        8  & Stockers and order fillers & 53-7065 & -1.210 & 8  & Childcare workers & 39-9011 & -1.443 \\
        9  & Cleaners of vehicles and equipment & 53-7061 & -1.199 & 9  & Waiters and waitresses & 35-3031 & -1.389 \\
        10 & Other entertainment attendants and related workers & 39-30XX & -1.164 & 10 & Manicurists and pedicurists & 39-5092 & -1.389 \\
        \bottomrule
    \end{tabular}%
    }
    \caption{Occupations in SW bin 1, ranked by lowest SW score.}
    \label{tab:US_sw_bin1_male_female}
\end{table*}

\begin{table*}[htbp]
    \centering
    \tiny
    \setlength{\tabcolsep}{2.5pt}    
    \resizebox{\textwidth}{!}{%
    \begin{tabular}{r p{6.6cm} c r @{\hspace{0.6cm}} r p{6.6cm} c r}
        \toprule
        \multicolumn{4}{c}{\textbf{Male-dominated occupations in SW bin 2}} &
        \multicolumn{4}{c}{\textbf{Female-dominated occupations in SW bin 2}} \\
        \multicolumn{4}{c}{\textit{Bin mean:} $-0.449$} &
        \multicolumn{4}{c}{\textit{Bin mean:} $-0.462$} \\
        \cmidrule(lr){1-4} \cmidrule(lr){5-8}
        Rank & Occupation & SOC Code & SW score & Rank & Occupation & SOC Code & SW score \\
        \midrule
        1  & Other woodworkers & 51-70XX & -0.462 & 1  & Billing and posting clerks & 43-3021 & -0.463 \\
        2  & Exercise trainers and group fitness instructors & 39-9031 & -0.433 & 2  & Postal service clerks & 43-5051 & -0.427 \\
        3  & Conveyor, dredge, and hoist and winch operators & 53-70XX & -0.472 & 3  & Title examiners, abstractors, and searchers & 23-2093 & -0.521 \\
        4  & Embalmers, crematory operators and funeral attendants & 39-40XX & -0.473 & 4  & Opticians, dispensing & 29-2081 & -0.392 \\
        5  & Agricultural inspectors & 45-2011 & -0.484 & 5  & Correspondence clerks & 43-4021 & -0.538 \\
        6  & Sailors and marine oilers & 53-5011 & -0.484 & 6  & Flight attendants & 53-2031 & -0.548 \\
        7  & Engine and other machine assemblers & 51-2031 & -0.488 & 7  & Public safety telecommunicators & 43-5031 & -0.573 \\
        8  & Chefs and head cooks & 35-1011 & -0.498 & 8  & New accounts clerks & 43-4141 & -0.342 \\
        9  & Structural metal fabricators and fitters & 51-2041 & -0.502 & 9  & Pharmacy technicians & 29-2052 & -0.590 \\
        10 & Solar photovoltaic installers & 47-2231 & -0.517 & 10 & Library technicians & 25-4031 & -0.592 \\
        \bottomrule
    \end{tabular}%
    }
    \caption{Occupations in SW bin 2, ranked by SW score closest to the bin mean.}
    \label{tab:US_sw_bin2_male_female}
\end{table*}

\begin{table*}[htbp]
    \centering
    \tiny
    \setlength{\tabcolsep}{2.5pt}    
    \resizebox{\textwidth}{!}{%
    \begin{tabular}{r p{6.6cm} c r @{\hspace{0.6cm}} r p{6.6cm} c r}
        \toprule
        \multicolumn{4}{c}{\textbf{Male-dominated occupations in SW bin 3}} &
        \multicolumn{4}{c}{\textbf{Female-dominated occupations in SW bin 3}} \\
        \multicolumn{4}{c}{\textit{Bin mean:} $0.279$} &
        \multicolumn{4}{c}{\textit{Bin mean:} $0.348$} \\
        \cmidrule(lr){1-4} \cmidrule(lr){5-8}
        Rank & Occupation & SOC Code & SW score & Rank & Occupation & SOC Code & SW score \\
        \midrule
        1  & Electrical power-line installers and repairers & 49-9051 & 0.270 & 1  & Respiratory therapists & 29-1126 & 0.340 \\
        2  & First-line supervisors of correctional officers & 33-1011 & 0.291 & 2  & Recreational therapists & 29-1125 & 0.338 \\
        3  & Tool and die makers & 51-4111 & 0.267 & 3  & Graphic designers & 27-1024 & 0.370 \\
        4  & Biological technicians & 19-4021 & 0.267 & 4  & Lodging managers & 11-9081 & 0.381 \\
        5  & Exercise physiologists & 29-1128 & 0.247 & 5  & Administrative services managers & 11-3012 & 0.384 \\
        6  & First-line supervisors of construction trades and extraction workers & 47-1011 & 0.242 & 6  & Probation officers and correctional treatment specialists & 21-1092 & 0.410 \\
        7  & Avionics technicians & 49-2091 & 0.237 & 7  & Social science research assistants & 19-4061 & 0.413 \\
        8  & Electrical and electronic engineering technologists and technicians & 17-3023 & 0.323 & 8  & Insurance sales agents & 41-3021 & 0.417 \\
        9  & First-line supervisors of mechanics, installers, and repairers & 49-1011 & 0.336 & 9  & Executive secretaries and executive administrative assistants & 43-6011 & 0.279 \\
        10 & Crane and tower operators & 53-7021 & 0.340 & 10 & Directors, religious activities and education & 21-2021 & 0.274 \\
        \bottomrule
    \end{tabular}%
    }
    \caption{Occupations in SW bin 3, ranked by SW score closest to the bin mean.}
    \label{tab:US_sw_bin3_male_female}
\end{table*}

\begin{table*}[htbp]
    \centering
    \tiny
    \setlength{\tabcolsep}{2.5pt}    
    \resizebox{\textwidth}{!}{%
    \begin{tabular}{r p{6.6cm} c r @{\hspace{0.6cm}} r p{6.6cm} c r}
        \toprule
        \multicolumn{4}{c}{\textbf{Male-dominated occupations in SW bin 4}} &
        \multicolumn{4}{c}{\textbf{Female-dominated occupations in SW bin 4}} \\
        \cmidrule(lr){1-4} \cmidrule(lr){5-8}
        Rank & Occupation & SOC Code & SW score & Rank & Occupation & SOC Code & SW score \\
        \midrule
        1  & Chief executives & 11-1011 & 2.414 & 1  & Nurse anesthetists & 29-1151 & 2.807 \\
        2  & Podiatrists & 29-1081 & 2.390 & 2  & Optometrists & 29-1041 & 2.118 \\
        3  & Architectural and engineering managers & 11-9041 & 2.332 & 3  & Pharmacists & 29-1051 & 1.972 \\
        4  & Lawyers & 23-1011 & 2.274 & 4  & Physician assistants & 29-1071 & 1.924 \\
        5  & Economists & 19-3011 & 2.145 & 5  & Veterinarians & 29-1131 & 1.888 \\
        6  & Petroleum engineers & 17-2171 & 1.976 & 6  & Nurse practitioners & 29-1171 & 1.835 \\
        7  & Mathematicians & 15-2021 & 1.864 & 7  & Clinical and counseling psychologists & 19-3033 & 1.542 \\
        8  & Computer and information research scientists & 15-1221 & 1.853 & 8  & Survey researchers & 19-3022 & 1.526 \\
        9  & Actuaries & 15-2011 & 1.798 & 9  & Audiologists & 29-1181 & 1.462 \\
        10 & Architects, except landscape and naval & 17-1011 & 1.669 & 10 & Physical therapists & 29-1123 & 1.343 \\
        \bottomrule
    \end{tabular}%
    }
    \caption{Occupations in SW bin 4, ranked by highest SW score.}
    \label{tab:US_sw_bin4_male_female}
\end{table*}

\section{Conclusions}
\label{sec:conclusions}

This paper investigates how different occupations are exposed to AI and how this may affect existing gender inequalities in the labour market, with particular attention to differences across the skill and wage distributions. This contribution is significant as it moves beyond existing studies that examine AI exposure in isolation and investigates whether AI exposure differs systematically across the skill and wage distributions within male- and female-dominated occupations. This enables us to assess how AI may affect the most vulnerable segments of the population, i.e., those in lower-skilled and lower-paid positions. Our findings show that AI exposure is not uniform. Within female-dominated occupations, similar levels of AI exposure are observed across both high-skilled, high-paid roles and low-skilled, low-paid roles. This contrasts with the pattern observed in male-dominated occupations, where higher AI exposure tends to be concentrated in more skilled and better-paid occupations. These results point to important forward-looking risks. In particular, triangulation with the existing literature suggests that women in more vulnerable positions (i.e., lower-skilled and lower-paid occupations) may be exposed to forms of AI that are more likely to replace workers rather than assist them, potentially increasing their risk of displacement and limiting career progression. Additionally, AI exposure differs depending on the type of AI considered. In particular, recent advances in LLMs appear to be more closely associated with female-dominated occupations, while broader AI innovations remain more concentrated in male-dominated occupations. These findings have policy significance, highlighting the need to target reskilling and labour-market support towards groups most at risk, thereby limiting the potential for AI to widen existing inequalities. While a number of limitations are acknowledged (see Section \ref{sec:discussion}), the analysis presented in this work is supported by the integration of complementary datasets, the use of two alternative measures of AI exposure to capture different dimensions of the phenomenon, and the triangulation of results against the existing literature.

\section* {Ethical Statement}

This study relies exclusively on publicly available, aggregated occupational data and does not involve the collection, processing, or analysis of any individual-level or personally identifiable information. The datasets used, including O*NET occupational descriptors, U.S. Census-based income distributions, and AI exposure indices derived from publicly available sources, are used solely for the purpose of aggregate-level labour market analysis.

The analysis focuses on occupational categories rather than individuals, and all results are reported at the level of occupations or occupational groups. As such, the study does not enable the identification of individuals and does not involve direct interaction with human subjects.

A key ethical consideration in this work relates to the interpretation and potential societal implications of AI exposure measures. While the indices used in this study capture potential exposure to AI-related technological change, they do not measure realised outcomes such as job displacement, wage changes, or productivity effects. Care has therefore been taken to avoid deterministic interpretations of exposure as direct evidence of harm or benefit.

Finally, we note that research on AI and labour markets has potential policy relevance. Care has been taken to ensure that results are communicated in a way that avoids reinforcing stereotypes or implying inherent differences in ability across gender groups. 

\textbf{Declaration of use of Generative AI.} The authors used Antigravity to assist with partial code generation, and Gemini, ChatGPT, and Grammarly to support grammar checking, rephrasing, and improvements to clarity. All analysis was conducted by the authors, who reviewed, verified, and edited all AI-assisted outputs.

 \section*{Acknowledgements}
 {This work was supported by the Engineering and Physical Sciences Research Council [grant number EP/Y009800/1], through funding from Responsible Ai UK (CC-00122).}

\bibliography{aaai2026}

\end{document}